\documentclass[%
 reprint,
 amsmath,amssymb,
 aps,
 prl,
 superscriptaddress
]{revtex4-2}

\usepackage{graphicx,xcolor}
\usepackage{dcolumn}
\usepackage{bm}
\usepackage[percent]{overpic}
\usepackage{physics}
\usepackage{mathtools}
\usepackage{hyperref}
\usepackage{booktabs}
\usepackage{longtable}
\usepackage{wasysym}
\usepackage{verbatim}

\begin{document}
\title{Gauge-theoretic origin of Rydberg quantum spin liquids}
\author{P. S. Tarabunga}
\affiliation{The Abdus Salam International Centre for Theoretical Physics (ICTP), strada Costiera 11, 34151 Trieste,
Italy}
\affiliation{International School for Advanced Studies (SISSA), via Bonomea 265, 34136 Trieste, Italy}
\affiliation{INFN, Sezione di Trieste, Via Valerio 2, 34127 Trieste, Italy}
\author{F. M. Surace}
\affiliation{Department of Physics and Institute for Quantum Information and Matter,
California Institute of Technology, Pasadena, California 91125, USA}
\author{R. Andreoni}
\affiliation{The Abdus Salam International Centre for Theoretical Physics (ICTP), strada Costiera 11, 34151 Trieste,
Italy}
\affiliation{International School for Advanced Studies (SISSA), via Bonomea 265, 34136 Trieste, Italy}
\affiliation{Dipartimento di Fisica “G. Occhialini”, Universit\`a degli Studi di Milano-Bicocca, Piazza della Scienza 3, I - 20126 Milano, Italy}
\author{A. Angelone}
\affiliation{Sorbonne Université, CNRS, Laboratoire de Physique Théorique de la Matière Condensée, LPTMC, F-75005 Paris, France}
\author{M. Dalmonte}
\affiliation{The Abdus Salam International Centre for Theoretical Physics (ICTP), strada Costiera 11, 34151 Trieste,
Italy}
\affiliation{International School for Advanced Studies (SISSA), via Bonomea 265, 34136 Trieste, Italy}

\date{\today}

\begin{abstract}
Recent atomic physics experiments and numerical works have reported complementary signatures of the emergence of a topological quantum spin liquid in 
models 
with blockade interactions. However, the specific mechanism  stabilizing such a phase remains unclear. 
Here, we introduce an exact relation between an Ising-Higgs 
lattice gauge theory on the kagome lattice and blockaded models on Ruby lattices. This relation elucidates the origin of previously observed topological spin liquids by directly linking the latter to a deconfined phase of a solvable  gauge theory. By means of exact diagonalization and unbiased quantum Monte Carlo simulations, we show that the
deconfined phases extend in a broad region of the parameter space; these states are characterized by a large ground state overlap with resonating valence bond wavefunctions. These blockaded models include both creation/annihilation and hopping dynamics, and can be experimentally realized with Rydberg-dressed atoms, offering novel and  controllable platforms for the engineering and characterisation of spin liquid states.

\end{abstract}
\maketitle

Introduced by Kitaev as a model for fault-tolerant quantum computation, the toric code is one of the most renowned examples of topological order \cite{KITAEV20032}. Its ground state, a $\mathbb{Z}_2$ quantum spin liquid (QSL) characterized by long-range entanglement, topological degeneracy and fractionalized excitations, has had a profound impact on both quantum information and condensed matter physics~\cite{Savary_2016,Wen2017}. In particular, it has shed considerable light onto the use of topological matter to encode quantum information in a robust manner. One of its key elements is its particularly clear connection to the gauge-theoretical origin of QSLs~\cite{moessner2001ising,lacroix2011introduction,fradkin2013field,moessner_moore_2021}: the presence of a $\mathbb{Z}_2$ gauge symmetry allows to interpret the latter as a deconfined phase of matter, where excitations (e.g., anyons) can be separated at arbitrary distance at finite energy cost, in sharp contrast to what happens in confined phases~\cite{wilson1974confinement,creutz1983quarks}.

Despite the aforementioned clear theoretical interpretation, experimentally observing such deconfined phases has proven challenging, mostly because directly engineering the toric code is highly non-trivial. A recent development has been the proposal that topological QSL can be realized in Rydberg atom arrays~\cite{browaeys2020many,kaufman2021quantum,Samajdar_2021}: in these systems, constraints akin to Gauss laws (or, equivalently, dimer constraints~\cite{lesanovsky2012interacting}) are imposed by means of the phenomenon of Rydberg blockade~\cite{lukin2001dipole,Glaetzle2014,Celi2020,glaetzle2015designing,surace2020lattice,Giudice2022}, that describes the incapability of exciting pairs of atoms within a certain radius. 

Recent numerical work has reported signatures of QSLs in a constrained model on the Ruby lattice, both at equilibrium~\cite{Verresen2021} and at the level of diabatic state preparation 
~\cite{giudici2022dynamical}: in the following, we will refer to this state as a Rydberg quantum spin liquid (RQSL). In a similar model~\cite{Semeghini2021}, 
strong signatures of deconfinement of a gauge theory have also been observed in experiments, albeit not yet in the ground state. Despite these remarkable results, the theoretical origin of such a deconfined phase remains so far unclear. While the similarity between the Rydberg blockade and a dimer constraint suggests that the RQSL phase can be understood as a resonant dimer state on the kagome lattice (which is known to be a $\mathbb{Z}_2$ QSL state \cite{Misguich_2002,Hao_2014}), this interpretation is challenged by the fact that no QSL phase is found in the dimer-model limit. In fact, signatures of a QSL are observed only when a moderate density of {\it monomers} is present: if this density is too low or too high the system will be in a crystalline or in a trivial phase, respectively. It remains an open question what is the mechanism that can allow for the emergence of topological order in such a small range of densities, with the additional crux that, at odds with gauge-theoretical expectations~\cite{fradkin1979phase,creutz1983quarks}, such a QSL appears in the absence of plaquette terms.

Here we address this conundrum by proving an exact mapping between $\mathbb{Z}_2$ lattice gauge theories \cite{Kogut1979} on the kagome lattice and a class of constrained models on the ruby lattice. Using this connection, we find a regime where topological order can be analytically established, showing how a resonating valence bond (RVB)~\cite{lacroix2011introduction} state is the {\it exact} ground state wavefunction of a Hamiltonian featuring solely two-body interactions. We then study the phase diagram beyond this regime using numerical simulations. Our results suggest that the QSL phase reported in Ref.~\cite{Verresen2021} is adiabatically connected to the exactly soluble point in an extended parameter space: importantly, according to our mapping, the experimentally relevant point is compatible with deconfinement, as its dual model features {\it strong} plaquette interactions - albeit, rather peculiarly, only on a certain sublattice. We then discuss how the type of models resulting from our mapping can be realized using Rydberg dressing techniques~\footnote{We note that our discussion specifically applies to the experimental setups inspired by Ref.~\cite{Semeghini2021}, and is complementary to other approaches to spin liquids in Rydberg systems~\cite{glaetzle2015designing,slagle2022quantum,Ohler2022,Weber2022}.}.

\begin{figure*}
    \centering
    \includegraphics[width=0.95\linewidth]{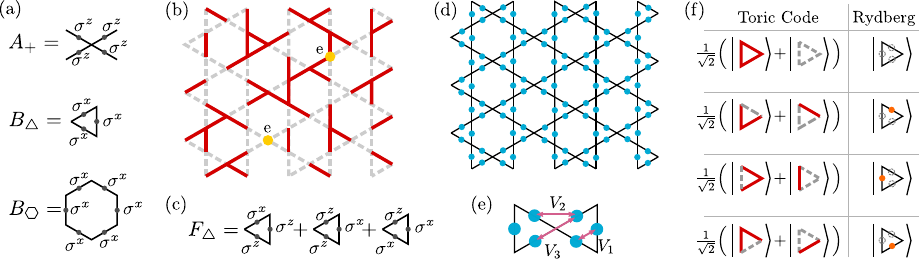}
    \caption{(a) Definition of vertex operators $A_+$ and plaquette operators $B_\triangle$, $B_{\hexagon}$. (b) Example of a configuration in the $\sigma_z$ basis: links marked with red solid (grey dashed) lines correspond to $\sigma^z=+1$ ($\sigma^z=-1$). Electric excitations are located on vertices with $A_+=1$. (c) Definition of the operator $F_\triangle$. (d) Examples of atom positions on the ruby lattice, formed by the midpoints on the links of a kagome lattice. (e) Definition of the interaction terms: nearest, next-nearest, and next-next-nearest neighbor interactions are $V_1 = \infty$, $V_2 = 4W$ and $V_3 = 4W$, respectively. Longer-range interactions are neglected. (f) Local mapping between the Hilbert spaces of a triangle in the toric code (left) and Rydberg atom models (right). The toric code space is constrained by $B_\triangle=1$. Solid red (grey dashed) lines indicate links with $\sigma^z=1$ ($\sigma^z=-1$). The Rydberg atom Hilbert space is constrained by the Rydberg blockade phenomenon: a triangle can have either 0 or 1 Rydberg excitations (orange circles).}
    \label{fig:fig1big}
\end{figure*}


\paragraph{Lattice gauge theory --}
\label{sec:tc}
We are interested in the toric code Hamiltonian on the kagome lattice, where Pauli operators $\sigma_j^\alpha$ are assigned to each link $j$. We define the operators $A_+$, $B_{\triangle}$ and $B_{\hexagon}$ on vertices, triangular, and hexagonal plaquettes, respectively, as (see Fig. \ref{fig:fig1big}-a):

\begin{equation}
    A_+=\prod_{j\in +} \sigma_j^z, \hspace{0.8cm}
    B_\triangle=\prod_{j\in \triangle} \sigma_j^x, \hspace{0.8cm}
    B_{\hexagon}=\prod_{j\in \hexagon} \sigma_j^x.
\end{equation}
$A_+,B_\triangle$ and $B_{\hexagon}$ form a mutually commuting set. Our Hamiltonian of interest has the form
\begin{equation}
\label{eq:tc}
    H_0^\text{TC} = W\sum_{+}A_+-J_1\sum_\triangle B_{\triangle}-J_2\sum_{\hexagon} B_{\hexagon}-g\sum_j \sigma_j^x.
\end{equation}
Let us first focus on the case $g=0$. Here, the ground state is obtained for $A_+=-1$ on every vertex and $B_{\triangle/\hexagon}=1$ on every plaquette, and is a (odd) $\mathbb{Z}_2$ spin liquid, with anyonic excitations. A vertex with $A_+=1$ represents an {\it electric} excitation or {\it charge} (see Fig. \ref{fig:fig1big}-b), with mass $2W$, while a plaquette with $B_{\triangle}=-1$ ($B_{\hexagon}=-1$) is a {\it magnetic} excitation or {\it vison}, with mass $2J_1$ ($2J_2$).

For $g\neq 0$ the Hamiltonian does still commute with $B_\triangle$ and $B_{\hexagon}$, but does not commute with the $A_+$ operators. The effect of the term proportional to $g$ is to excite pairs of electric charges on neighbouring sites; the model is a Ising-Higgs lattice gauge theory \cite{fradkin1979phase,Beekman_2017}, as it describes a $\mathbb{Z}_2$ gauge field coupled to matter. Since the toric-code ground state is gapped, for sufficiently small $g$ the system remains a $\mathbb{Z}_2$ spin liquid~\cite{fradkin1979phase,moessner_moore_2021}. For $g/W$ larger than a critical value, the model has a phase transition to a confined phase (a condensate of electric charges). Utilizing 
Wegner duality~\cite{Wegner}, we quantitatively establish the extent of the topological phase until $g_c/W=\widetilde g_c\simeq 0.3387$~\cite{supmat}. This argument and the location of the phase transition are valid for arbitrary $J_1$ and $J_2$, as long as the ground state is in the sector with no magnetic excitations. The model in Eq. (\ref{eq:tc}) can equivalently be interpreted as a pure Ising gauge theory on the dual lattice (a {\it dice lattice}), where electric and magnetic charges are interchanged: the term proportional to $g$ now excite pairs of magnetic excitations on neighboring plaquettes while the electric charge is static, the plaquette terms $B_{\triangle,\hexagon}$ become vertex terms, and Gauss' law is enforced on every vertex.

In our mapping to the Rydberg model below, we will consider  $J_1\rightarrow\infty$ and $J_2=0$: in this case, we are restricted to the sector with $B_\triangle=1$ on every triangular plaquette, while hexagonal plaquettes have zero ''mass" for $g=0$. While the absence of a gap at $g=0$ may be concerning, as it does not guarantee that the system is in a gapped spin liquid state in a finite range of $g/W$, we will show via both numerical simulations and perturbation theory that the ground state is always in the sector with no magnetic excitations and that a gapped spin liquid is present in a region with $0<g/W<\widetilde g_c$. This is because the term proportional to $g$ generates an effective mass for magnetic excitations, allowing for a gapped ground state with topological order even for $J_2=0$.

\paragraph{Rydberg model --}


We now map the toric code on the kagome lattice to a model of Rydberg atoms constrained by the Rydberg blockade. The atoms are located on the links of the lattice, forming a ruby lattice (see Fig. \ref{fig:fig1big}-d). The Rydberg excitations are modeled as hard-core bosons with operators $b_j, b_j^\dagger$. We assume that interactions between neighbouring atoms (i.e., belonging to the same triangle) are repulsive and very strong, inducing a nearest-neighbor blockade. Under this assumption, each triangle can have either $0$ or $1$ Rydberg excitation: each triangle has a Hilbert space spanned by the four basis states in Fig. \ref{fig:fig1big}-f (right column). Note that this assumption differs from the one of Ref. \cite{Verresen2021}, where the blockade extends to next-nearest and next-next-nearest neighbours: this naturally leads to different gauge constraints.

In the model in Eq.~(\ref{eq:tc}), the constraint $B_\triangle=1$ reduces the Hilbert space dimension of a triangle from $2^3$ to $4$. The constrained Hilbert space is spanned by the states in Fig. \ref{fig:fig1big}-f (left column),  defining a local mapping between the Hilbert spaces of the two models. This correspondence allows a direct mapping between the Hamiltonian in Eq. (\ref{eq:tc}) with $J_1\rightarrow \infty$, $J_2=0$ and the following Hamiltonian for Rydberg excitations \cite{supmat}:
\begin{multline}
\label{eq:Ryd}
    H^\text{Ryd}_0=-g\sum_j (b_j+b_j^\dagger)-g\sum_{\langle i,j\rangle}(b_i^\dagger b_j+b_j^\dagger b_i)\\
    -4W\sum_j n_j +4W\sum_{\langle \langle i,j \rangle \rangle} n_i n_j +4W\sum_{\langle \langle \langle i,j \rangle \rangle \rangle}  n_i n_j.
\end{multline}
Here, the first term is the creation/annihilation of Rydberg excitations, the second term is the hopping of excitations between neighbouring atoms, the third term is a chemical potential for the excitations, and the fourth and fifth terms are Rydberg interactions between next-nearest neighbors and next-next-nearest neighbors respectively.
Note that, while the nearest-neighbour interaction is $V_1=\infty$, next-nearest-neighbor and next-next-nearest-neighbor interactions are $V_2=V_3=4W$ (see Fig. \ref{fig:fig1big}-e). 
The experimental realization of this Hamiltonian will be further discussed below. 

A first result of the mapping above is that 't Hooft and Wilson lines in the toric code model are mapped onto the string operators proposed in Ref. \cite{Verresen2021}~\footnote{We nevertheless stress that utilizing such operators as order parameters for deconfinement should be done with great care: the correct order parameter~\cite{fredenhagen1986confinement} is defined by correlations in real space and imaginary time, and it is not necessarily matching real space correlations in lattice models with very small correlation length (for a spectacular failure, see Ref.~\cite{tschirsich2019phase}). While our work suggests a very strong correspondence between those and deconfinement even in real space, a quantitative verification of their regime of applicability is an interesting, open question.}: this suggests the QSL phase observed there has the same nature as the topological phase of the model in Eq. (\ref{eq:tc}), that is, stemming from a genuine $\mathbb{Z}_2$ gauge theory. Moreover, the resonating valence bond (RVB) state of dimers in \cite{Verresen2021} is here mapped to the toric code ground state (equal weight superposition of the configurations with $A_+=-1$ on each vertex): our mapping provides a local unitary transformation that relates the RVB dimer state and the toric code ground state; the existence of such a unitary was proven in \cite{schuch2012}. 

\paragraph{Additional terms --}
In order to emphasize the wide applicability of our reasoning in terms of experimental platforms, we now consider a broader class of dynamics, introducing additional Hamiltonian terms:
\begin{equation}
\label{eq_ryd1}
    H^\text{Ryd}_1=
    -h\sum_j (b_j+b_j^\dagger)+h\sum_{\langle ij\rangle}(b_i^\dagger b_j+b_j^\dagger b_i)\\
    -4\lambda\sum_j n_j,
\end{equation}
which, under the local mapping in \ref{fig:fig1big}-f become
\begin{equation}
    H_1^{TC}=
    -h\sum_\triangle F_{\triangle}+\lambda\sum_{\langle i,j \rangle} \sigma_i^z \sigma_j^z,
\end{equation}
where $F_\triangle$ is the operator in Fig. \ref{fig:fig1big}-c. Note that $H_1^{TC}$ commutes with the triangular plaquette operators $B_\triangle$, but does not commute with the hexagonal plaquette operators $B_{\hexagon}$. In particular, the term proportional to $h$ creates a pair of electric charges and a pair of visons (on hexagonal plaquettes) on the toric code ground state, while the term proportional to $\lambda$ creates pairs of visons on neighboring hexagons (i.e., hexagons that share a vertex). As shown in \cite{supmat}, 
the Hamiltonian $H^\text{TC}=H_0^\text{TC}+H_1^\text{TC}$ (and, equivalently, $H^\text{Ryd}=H^\text{Ryd}_0+H^\text{Ryd}_1$) has a self-duality transformation which maps $g\rightarrow -h$, $h\rightarrow -g$.

We remark that, by using the mapping defined for the states in Fig. \ref{fig:fig1big}-f, {\it any} Hamiltonian term of the Rydberg-atom model that is compatible with nearest-neighbor blockade (e.g., longer-range interactions) can be mapped to a Hamiltonian term of the toric code model. The blockade is the fundamental ingredient to enforce Gauss law~\cite{Glaetzle2014}.

\paragraph{Phase diagram --}

We now study the phase diagram of the Hamiltonian $H^\text{TC}=H_0^\text{TC}+H_1^\text{TC}$ by means of exact diagonalization, imposing the constraint $J_1\rightarrow \infty$ by directly restricting the Hilbert space to the states with $B_\triangle=1$ on every triangular plaquette. All results shown are for a 3x2 cluster containing 36 spins. Our goal here is not to precisely determine transition points, but rather, to discuss the generic structure of the phase diagram against theory. 

We first focus on the case $\lambda = 0$. For $g = 0$ or $h = 0$, the model can be mapped to the quantum Ising model on the kagome lattice \cite{supmat}, 
whose phase diagram has been obtained in previous studies \cite{Blote2002, Powalski2013,moessner2001ising}. To detect the transition points and obtain the phase diagram for arbitrary $g$, $h$, we compute the fidelity susceptibility $\chi_F$ defined as
\begin{equation}
    \chi_F(\eta)= \lim_{\delta\eta \rightarrow 0} \frac{-2\ln F(\eta,\delta\eta)}{(\delta\eta)^2}
\end{equation}
where the fidelity $F$ is defined as $F(\eta,\delta \eta)=|\langle \psi(\eta) | \psi(\eta+\delta\eta) \rangle|$, with $\eta$ being any parameter of the Hamiltonian. In systems with finite volume $L^d$,  $\chi_F/L^d$ is known to exhibit peaks, whose amplitudes diverge in the thermodynamic limit and whose positions converge towards the critical point; the position of the latter can be derived via finite-size scaling techniques \cite{Gu2009}.

\begin{figure}
    \centering
    \includegraphics[width=0.52\linewidth]{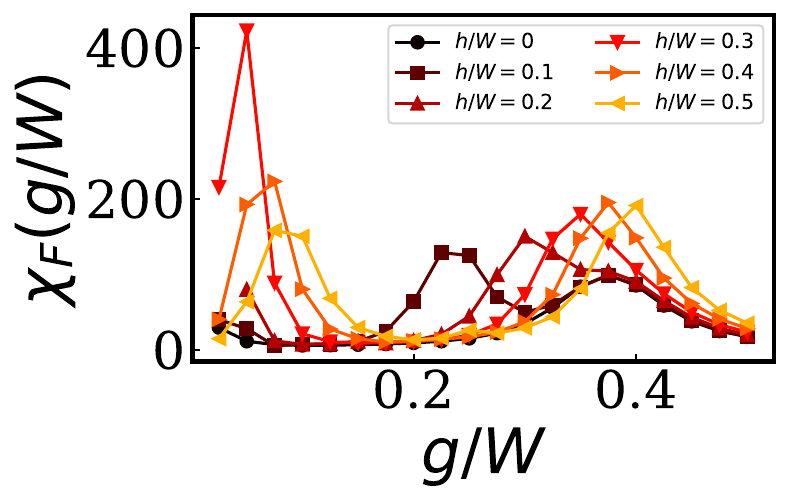}
    \includegraphics[width=0.46\linewidth]{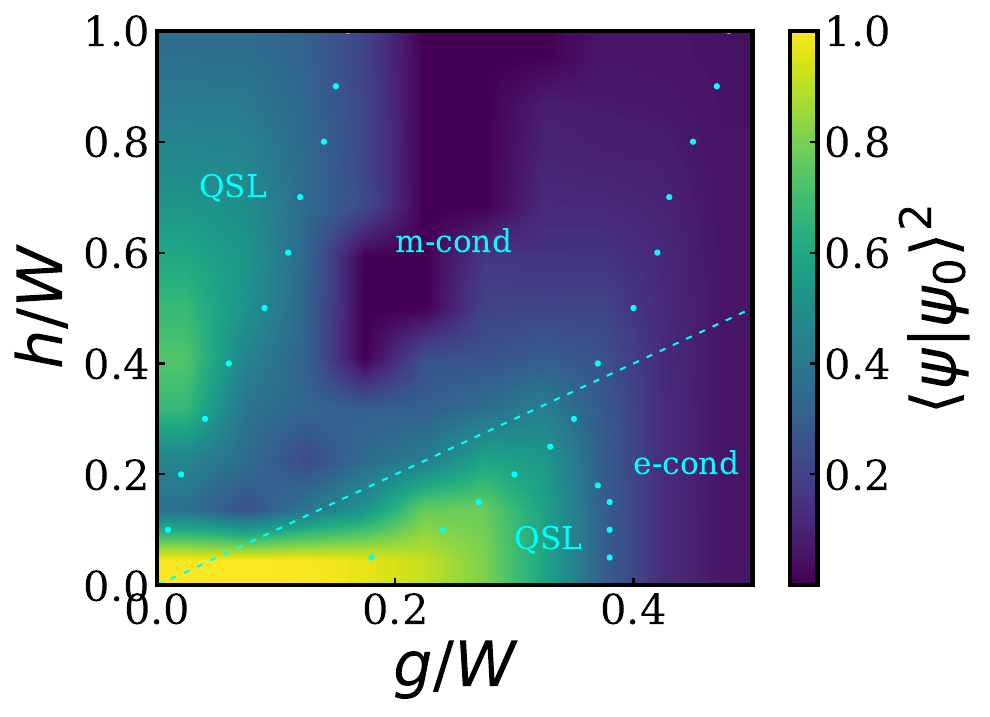}
    \caption{(a) Fidelity susceptibility of the Hamiltonian $H^\text{TC}$ for $\lambda=0$ as a function of the parameter $g$. (b) Phase diagram of $H^\text{TC}$ for $\lambda = 0$. The heatmap indicates the overlap of the ground state with $\ket{\psi_0}$. The dots indicate the peaks of the fidelity susceptibility.}
    \label{fig:lambda0}
\end{figure}

Fig. \ref{fig:lambda0}-a shows the fidelity susceptibility as a function of $g$ for $\lambda = 0$ (see \cite{supmat} 
for varying $h$). The marked dots in Fig. \ref{fig:lambda0}-b indicate the phase boundary, estimated as the position of the maximum of the fidelity susceptibility.
 
In addition, we compute the overlap of the ground state $\ket{\psi}$ with the state $\ket{\psi_0}$, defined as the ground state of the model for $h=0$, $\lambda=0$, $g/W=0.1$.
Since we know that the ground state of the model with $h = 0$, $\lambda = 0$ and small $g/W$ is a
QSL, the overlap of $\ket{\psi_0}$ with the ground state of the other points in the parameter space may tell us if those ground states have the same correlation structure. Fig. \ref{fig:lambda0}-b shows that, as expected, the overlaps are large in the QSL phases, and are close to 0 for the electric and magnetic condensates.

 \begin{figure}[t]
     \centering
    \includegraphics[width=0.99\linewidth]{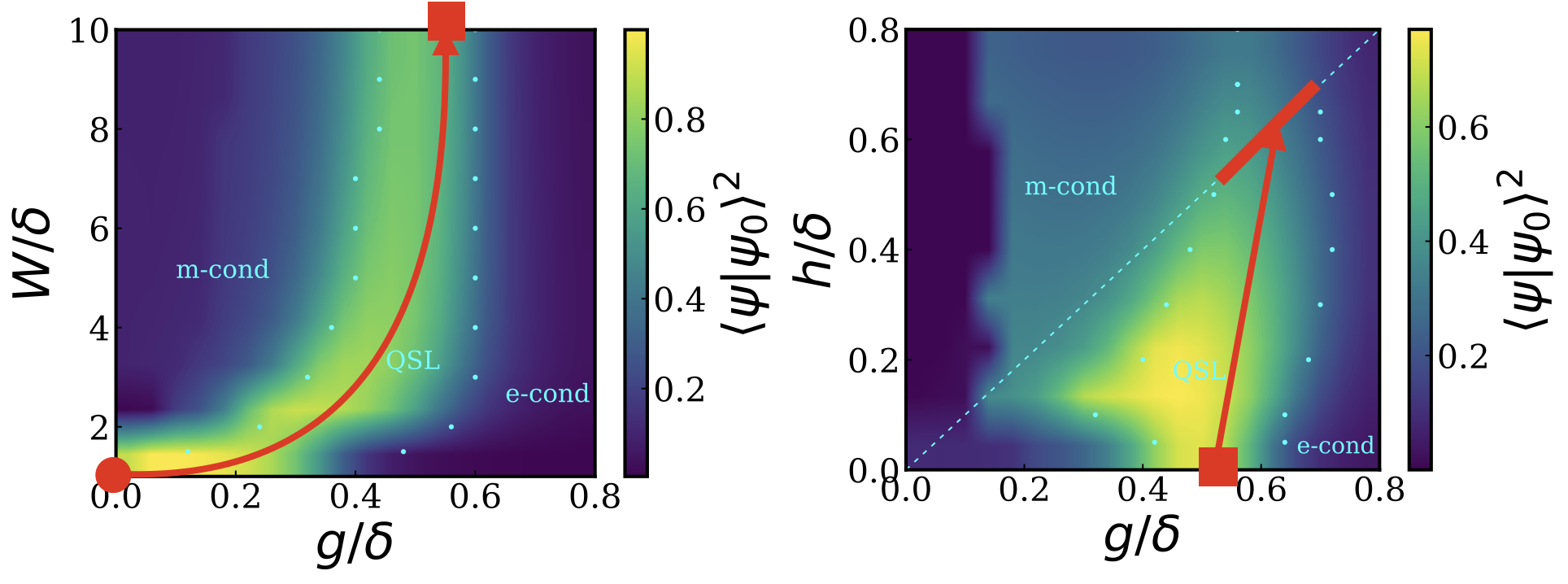}
     \caption{Adiabatic path from exactly soluble point (red dots) to the Rydberg line (thick red line): the square dot is common to both parts of the path. The overlap of the ground state of $H^\text{TC}$ with $\ket{\psi_0}$ at (a) $h = 0$ and (b) $W/\delta = 10$.}
     \label{fig:gh}
 \end{figure}
 
{\it Adiabatic path from exactly soluble point to the Rydberg line. -} A key question is, whether the exactly soluble point is adiabatically connected to the regime studied in Ref. \cite{Verresen2021}. To reproduce the parameter ranges considered there, we first define $\delta=W + \lambda$, and set it to be finite, while $W$ is taken to be large ($W = 10\delta$). 
We proceed in two steps: (i) on the $h=0$ hyperplane, we ramp $W/\delta$ from 1 to 10, and (ii) on the hyperplane $W=10\delta$, we ramp $h$ from 0 to the Rydberg line, $g=h$~\cite{Verresen2021}.

Fig.~\ref{fig:gh} shows the overlap with the exact RVB wave function in these parameter regimes; the various phases are separated by local peaks of the fidelity susceptibility~\cite{supmat}, indicated by dots. The red line is a sample of an adiabatic path entirely in the QSL phase: starting from the exactly soluble point (red circle), to the Rydberg line (thick red line in the right panel). As a non-trivial check on the latter line, we find two transitions at $g/\delta \approx 0.525$ and $g/\delta \approx 0.7$, corresponding to $\delta/\Omega = 1.4$ and $\delta/\Omega = 1.9$ in the notation of Ref. \cite{Verresen2021}, and compatible with the transitions found there.
Overall, these results show that the QSL phase observed in Ref. \cite{Verresen2021} is connected with the QSL phase that we demonstrated for $W/\delta = 1$.

\begin{figure}[b]
     \centering
     \includegraphics[width=0.85\linewidth]{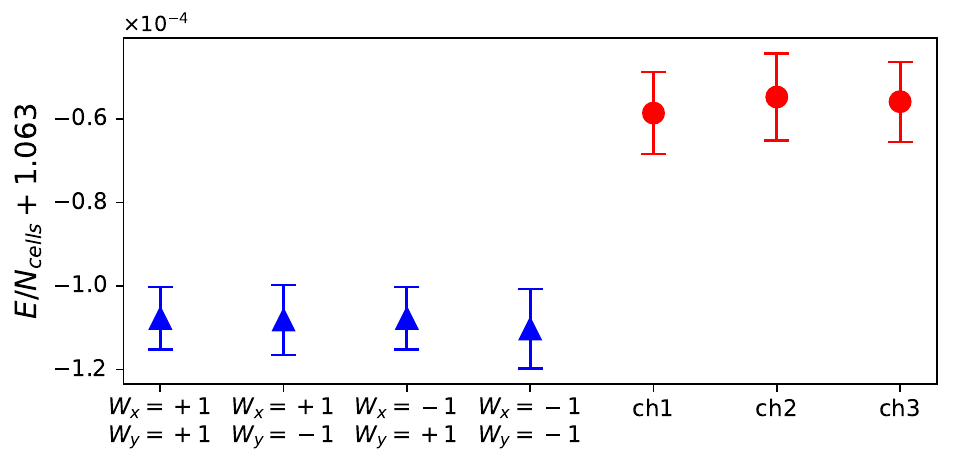}
     \caption{Rescaled energy per elementary cell of the quasi-degenerate ground states (blue triangles) and of the charged states (red circles) for the model $H^\text{TC}$ on a $N_{\mathrm{cell}} = 8 \times 8$ cluster and $g/W = 0.32$. All charged states displayed here are characterized by the Wilson loop signs $W_x = W_y = +1$. Errorbars correspond to a $95\%$ confidence interval.}
     \label{fig:spectrum}
 \end{figure}

\paragraph{Quantum Monte Carlo spectroscopy. --} It is expected that, within the entire topological phase, the ground state is approximately four-fold degenerate on torus geometries. An important question is how large is the energy gap separating the ground state manifold from excited states. This question is particularly relevant in this case, as there is no ``bare mass" for the magnetic excitations on hexagonal plaquettes (i.e., $J_2=0$). Along the line $h = 0$, we have devised a spectroscopy based on simulating, via unbiased Quantum Monte Carlo approaches~\cite{Sandvik1991, Sandvik2003}, a dual Ising model description of low-lying states (see~\cite{supmat}). A sample of the corresponding results are presented in Fig.~\ref{fig:spectrum} for $g/W=0.32$. The four-fold degeneracy of the ground state manifold is clear: the (expected) small splitting between those states is within our error bars. The gaps to the first excited states with magnetic excitations (labelled as ''ch1,2,3") are finite; their relatively small magnitude we attribute to the strong anisotropy of magnetic and electric excitations. 

\label{sec:exp}

\paragraph{Experimental realization --} The class of models in $H^{\text{Ryd}}$ considerably extends the range of target Hamiltonians that can realize a QSL. The salient features of Eq.~\eqref{eq:Ryd} and Eq.~\eqref{eq_ryd1} are naturally realized in arrays of atoms weakly laser-coupled to Rydberg states~\cite{Henkel2010,Pupillo2010,Macri2014,low2012experimental,glaetzle2015designing,Jau2016, Zeiher2016, Guardado-Sanchez2021, Mattioli2013, Cinti2014, Dalmonte2015, Angelone2016, Masella2019, Angelone2020,johnson2010,Zeiher_2017,Borish2020}, and trapped into very deep potentials (e.g., optical lattices or tweezers), so that tunneling is suppressed over the time scales of the experiment. 
We briefly discuss here the $h\neq g$ regime, and discuss an alternative implementation working for $h=g$ in Ref.~\cite{supmat}. 

We consider two states of an atom with different $m_F$, both belonging to the ground state manifold. They are laser-coupled to a Rydberg manifold with Rabi frequency $\Omega$, and detuning $\delta$, with $|\Omega|\ll\delta$. Within perturbation theory, the most relevant terms are: (a) a soft-shoulder potential that decays quickly to 0 after a cut-off distance that can be tuned by changing $\Omega$ and $\delta$: this realizes our potentials in the limit $g/W\ll 1$; and (b) a hopping term, induced by the off-diagonal part of the Rydberg interaction. The range of the latter can be tuned independently of the previous one~\cite{glaetzle2015designing}. The linear terms in $b,b^\dagger$ as well as the $\lambda$ term can be realized, e.g., utilizing a resonant microwave coupling.

\paragraph{Conclusions --}
In this work, we showed that a $\mathbb{Z}_2$ lattice gauge theory with matter on a kagome lattice can be exactly mapped to a blockaded model of atoms on the ruby lattice, realizable with Rydberg-dressed potentials. Building on this analytical finding, we proved the existence of a $\mathbb{Z}_2$ QSL and precisely located the transition to a trivial phase. By including additional perturbations, we extended our model to comprise, as a specific case, the Hamiltonian studied in Ref.\cite{Verresen2021,Semeghini2021}: our numerical results suggest that the QSL observed therein belongs to the same phase of our solvable case. By elucidating the origin of QSL phases in blockaded systems, our work opens interesting directions for future experiments: more robust QSL states could be realized in an enlarged class of models, and the gauge-theoretical interpretation can help to identify useful perturbations that stabilize QSL phases, as well as suitable observables to detect and characterize their topological properties.

\begin{acknowledgments}
We thank F. Becca, A. Browaeys, G. Giudici and O. Motrunich for insightful discussions, and M. Lukin for correspondence.
The work of A.A., M. D., P. S. T. and F.S. was partly supported by the ERC under grant number 758329 (AGEnTh), by the MIUR Programme FARE (MEPH), and by the European Union's Horizon 2020 research and innovation programme under grant agreement No 817482 (Pasquans). P.S.T acknowledges support from the Simons Foundation through Award 284558FY19 to the ICTP.

{\it Note added:} while this manuscript was in preparation, a preprint appeared proposing a parton mean-field description of Rydberg quantum spin liquids~\cite{samajdar2022emergent}. Moreover, a work appeared after our manuscript were posted on arxiv (2205.15302), that also shows adiabatic continuity between the Rydberg regime and an RVB state. 

\end{acknowledgments}
\bibliography{biblio}

\end{document}


\title{Supplemental material\\ Gauge-theoretic origin of Rydberg quantum spin liquids}
\author{Poetri Sonya Tarabunga}
\affiliation{The Abdus Salam International Centre for Theoretical Physics (ICTP), Strada Costiera 11, 34151 Trieste,
Italy}
\affiliation{International School for Advanced Studies (SISSA), via Bonomea 265, 34136 Trieste, Italy}
\affiliation{INFN, Sezione di Trieste, Via Valerio 2, 34127 Trieste, Italy}
\author{Federica M. Surace}
\affiliation{Department of Physics and Institute for Quantum Information and Matter,
California Institute of Technology, Pasadena, California 91125, USA}
\author{Riccardo Andreoni}
\affiliation{The Abdus Salam International Centre for Theoretical Physics (ICTP), strada Costiera 11, 34151 Trieste,
Italy}
\affiliation{International School for Advanced Studies (SISSA), via Bonomea 265, 34136 Trieste, Italy}
\affiliation{Dipartimento di Fisica “G. Occhialini”, Universit\`a degli Studi di Milano-Bicocca, Piazza della Scienza 3, I - 20126 Milano, Italy}
\author{Adriano Angelone}
\affiliation{Sorbonne Université, CNRS, Laboratoire de Physique Théorique de la Matière Condensée, LPTMC, F-75005 Paris, France}
\author{Marcello Dalmonte}
\affiliation{The Abdus Salam International Centre for Theoretical Physics (ICTP), strada Costiera 11, 34151 Trieste,
Italy}
\affiliation{International School for Advanced Studies (SISSA), via Bonomea 265, 34136 Trieste, Italy}
\date{\today}

\maketitle

\setcounter{equation}{0}
\setcounter{figure}{0}
\setcounter{table}{0}
\setcounter{page}{1}
\makeatletter
\renewcommand{\theequation}{S\arabic{equation}}
\renewcommand{\thefigure}{S\arabic{figure}}
\renewcommand{\bibnumfmt}[1]{[S#1]}
\renewcommand{\citenumfont}[1]{S#1}

Here we derive in detail some of the results reported in the main text, and present additional numerical data. This supplemental material is structured as follows: in Sec.
\ref{sec:mapping} we explicitly report the mapping of operators between the Ising-Higgs lattice gauge theory and Rydberg-atom model introduced in the main text; in Sec. \ref{sec:selfduality} we show that the model is self-dual; in Sec. \ref{sec:ising} we show how the Hamiltonian $H_0^\text{TC}$  projected to the different superselection sectors can be mapped to a class of Ising models; in Sec. \ref{sec:perturb} we compute the magnetic mass for small $g/W$ using perturbation theory; in Sec. \ref{sec:num} we report additional numerical results; finally, in Sec. \ref{sec:altexp} we propose an alternative experimental implementation of the model along the $g=h$ line.

\section{Mapping}
\label{sec:mapping}
In this section, we show the mapping of $H^{TC}$ to $H^\text{Ryd}$. First of all,  since all triangular plaquettes are independent, the Hilbert space can be decomposed as $\mathscr{H} = \bigotimes_{\triangle} \mathscr{H}_{\triangle}$. Both the constraints $B_{\triangle}$ and $n_in_j=0$ (for nearest neighbors $i,j$) restrict the Hilbert space of a single triangle from 8 to 4 dimensions. We take a basis of 4 states of a triangle in the $\mathbb{Z}_2$ LGT model and map them to 4 states of a triangle in the Rydberg model, as shown in Fig.1-f.

With this mapping, the Hamiltonian terms are transformed in the following way,
\begin{equation} \label{eq:mapping}
\begin{split}
    \sigma_i^x\left( \frac{1 + \sigma_{j}^z\sigma_{k}^z}{2} \right)
    &\longmapsto 
    (b^\dagger_i +b_i) \\
    \sigma_k^x\left( \frac{1 - \sigma_{i}^z\sigma_{j}^z}{2} \right)
    &\longmapsto 
    (b_i b^\dagger_j +h.c.) \\
    \prod_{i\in\times} \sigma^z_i &\longmapsto \prod_{i\in{\large\times}} (2n_i - 1) \\
    \sum_{\langle ij \rangle \in \triangle} \sigma^z_i\sigma^z_j &\longmapsto  -4\sum_{j\in\triangle} n_j + 3
\end{split}
\end{equation}
where $i,j$ and $k$ are the three links belonging to a single triangle. We note that terms of the form $b^\dagger_i b^\dagger_j$ are not present in the model (and, in the implementation discussed in the main text, can be systematically avoided as discussed in Ref.~\cite{glaetzle2015designing}).

To simplify the term $\prod_{j\in{\large\times}} (2n_j - 1)$, consider the links $i,j,k,l\in\times$ with $i,j$  belonging to the same triangle and $k,l$ belonging to a different triangle. Then,
\begin{equation} 
\begin{split}
   \prod_{j\in{\large\times}} (2n_j - 1)&=(2n_i-1)(2n_j-1)(2n_k-1)(2n_l-1) \\
   &=(4n_in_j-2n_i-2n_j+1)(4n_kn_l-2n_k-2n_l+1)\\
   &=(-2n_i-2n_j+1)(-2n_k-2n_l+1)\\
   &=4n_in_k+4n_in_l+4n_jn_k+4n_l-2(n_i+n_j+n_k+n_l)+1.
 \end{split}
\end{equation}
Hence, we obtain
\begin{equation} \label{eq:mapping2}
    \prod_{j\in{\large\times}}\sigma_j^z
    \longmapsto 
    1 -2 \sum_{j\in\large\times}n_j +4 \sum_{i\in>}\sum_{j\in<}n_in_j
\end{equation}
Note that each link belongs to two vertices, hence
\begin{equation}
    \sum_{\large\times}\sum_{j\in\large\times}n_j = 2\sum_{j}n_j
\end{equation}
Moreover, $\sum_{i\in>}\sum_{j\in<}n_in_j$ contains four pairs: two at distance $r_2$ and two at distance $r_3$, i.e., they are respectively second and third nearest neighbors, $\langle\langle i,j \rangle\rangle$ and $\langle\langle\langle i,j \rangle\rangle\rangle$. Finally, summing over all triangles and vertices, the mappings in Eq. (\ref{eq:mapping}) and (\ref{eq:mapping2}) define a mapping (up to constants) of $H^\text{TC}$ to $H^\text{Ryd}$.

\section{Self-duality}
\label{sec:selfduality}
We now show that the Hamiltonian $H_0^\text{TC}+H_1^{TC}$ with $J_2=0$ has a self-duality that maps $g\rightarrow -h$, $h\rightarrow -g$. The transformation we consider is induced by the following unitary mapping:
\begin{equation}
\label{eq:unitary}
    U= \exp\left[-i\frac{\pi}{4}\sum_\triangle\left(\sum_{j \in \triangle} \sigma_j^z+\prod_{j\in \triangle}\sigma_{j}^z \right)\right]=\prod_\triangle \left[\frac{1}{2} \left(1-\sum_{\langle i,j \rangle\in \triangle} \sigma_i^z \sigma_j^z\right)\right]
\end{equation}
We remark that, in contrast with other dualities of the theory, this transformation has a completely local nature.
The action of this transformation on the Pauli operators of the link $j$ is
\begin{align}
    U\sigma_j^zU^\dagger =&\sigma_j^z,\\
    U\sigma_j^x U^\dagger =& -\sigma_j^x\sigma_i^z\sigma_k^z,
\end{align}
where $i$ and $k$ are the two links that belong to the same triangle as $j$. It can be easily checked that $UB_\triangle U^\dagger=B_\triangle$ and $UA_+U^\dagger=A_+$, while
\begin{align}
    U \left(\sum_{j \in \triangle}\sigma_j^x\right)U^\dagger =-F_{\triangle},\\
    U F_{\triangle} U^\dagger =-\sum_{j \in \triangle}\sigma_j^x.
\end{align}
Therefore, the transformation sends $g\rightarrow -h$, $h\rightarrow -g$ while leaving $W$, $J_1$ and $\lambda$ invariant. Note that, while $B_\triangle$ and $A_+$ are invariant under the transformation, the hexagonal plaquette $B_{\hexagon}$ is not, as it gets decorated by $\sigma^z$ operators.

Under the mapping to the Rydberg model, the unitary $U$ in Eq. (\ref{eq:unitary}) takes the form
\begin{equation}
    U=\prod_j (-1)^{n_j},
\end{equation}
and therefore induces the transformations
$Ub_j U^\dagger= -b_j,\; Ub^\dagger_j U^\dagger= -b^\dagger_j,\; Un_j U^\dagger= n_j$.

A consequence of the self-duality illustrated here is that the phase diagram in the $g-h$ plane is symmetric under reflection with respect to the axis $h=-g$, and that the line $g=0$ for $\lambda=0$ can also be solved analytically with a mapping to the Ising model, similarly to the $h=0$, $\lambda=0$ case. On the self-dual line $h=-g$ the quantity $\sum_{\langle i,j\rangle}\sigma_i^z\sigma_j^z$ (and, equivalently, the total number of Rydberg excitations $\sum_j n_j$) is conserved.

\section{Wegner duality and Ising model}
\begin{figure}[h]
    \centering
    \includegraphics[width=\linewidth]{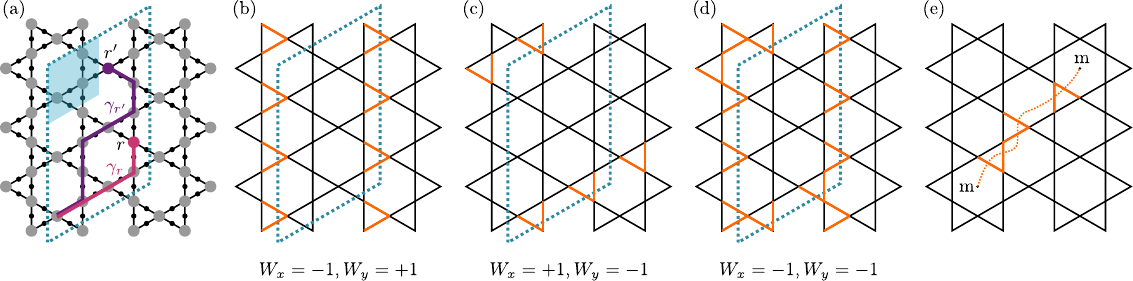}
    \caption{(a) The LGT variables $\sigma_j^\alpha$ are located on the links of the kagome lattice (black dots). The Ising variables are defined on the vertices of the kagome (grey circles). A unit cell (blue shaded area) contains 6 LGT sites, 3 Ising sites, 2 triangular plaquettes, 1 hexagonal plaquette. The dashed blue line delimits a system with $N_c=6$ cells. Periodic boundary conditions are imposed. The paths $\gamma_r, \gamma_{r'}$ connect a designated vertex (here on the left-bottom corner) with sites $r$ and $r'$. (b-d) Sectors with different values of the Wilson loops are obtained by setting $s_j=-1$ on the links crossing the horizontal/vertical boundaries. The links in orange are the ones with $s_j=-1$. (e) A sector with a pair of magnetic excitations can be obtained with an appropriate choice of signs of $s_j$: a line is drawn that connects the two magnetic excitations; for all the links crossed by the line we set $s_j=-1$. This choice ensures that $B_{\hexagon}=-1$ for the two plaquettes with the excitations, and $B_{\triangle}=1$, $B_{\hexagon}=1$ for all the other plaquettes.}
    \label{fig:wegner}
\end{figure}

 In this section, we show the mapping of the Hamiltonian $H_0^\text{TC}$ to the quantum Ising model on the kagome lattice. This mapping follows the general approach of Wegner dualities \cite{Wegner}.
 Although the hexagonal plaquette operators $B_{\hexagon}$ do not appear in the Hamiltonian, they do commute with the Hamiltonian. Therefore, we can study the  model separately in each sector of fixed eigenvalues of $B_{\hexagon}$ and the Wilson loops $W_{x,y}$. Each sector can be mapped to a transverse-field Ising model on the kagome lattice with appropriate nearest-neighbor coupling signs. To see this, we define an Ising variable on each vertex $r$ of a kagome lattice
\begin{equation} \label{eq:ising_var}
\begin{split}
    \tau^z_r&=\prod_{j\in\times} \sigma^z_j \\
    \tau^x_r&=\prod_{j\in\gamma_r} s_j \sigma^x_j,
\end{split}
\end{equation}
where the signs $s_{r,r'}=\pm1$ are chosen such that $\prod_{j\in \triangle,\hexagon}s_j=B_{\triangle,\hexagon}$ and $\prod_{j\in \gamma_{x,y}}s_j=W_{x,y},$ and $\gamma_r$ is a path between a specific vertex and $r$ (see Fig. \ref{fig:wegner}-a). Note that the variables $\tau^x_r$ are defined independently of the choice of the path $\gamma_r$ because $\prod_{j\in\Gamma} s_j \sigma^x_j=1$ for every closed loop $\Gamma$.  The variable $\tau^z_r$ indicates the presence of/absence of an electric charge in $r$. 

With the mapping in Eq. (\ref{eq:ising_var}), the toric code Hamiltonian is mapped to
\begin{equation} \label{eq:ising}
    H=-g\sum_{\langle r,r' \rangle} s_{r,r'} \tau_r^x \tau_{r'}^x + W \sum_r \tau_r^z.
\end{equation}


A few comments are in order. In the original lattice gauge theory, the number of degrees of freedom once the plaquettes and the Wilson loops are fixed is $6N_c-N_p-N_W+1=3N_c-1$, where $N_c$ is the number of unit cells in the system (each unit cell has 6 spins, see Fig. \ref{fig:wegner}-a), $N_p=3N_c$ is the number of plaquettes, $N_W=2$ is the number of independent Wilson loops, and we also take into account the fact that not all plaquettes are independent, because the product of all plaquettes is $1$. In the corresponding Ising model, each unit cell contains $3$ Ising variables, so the total number of degrees of freedom is $3N_c$. To obtain the correct counting, we have to further fix the value of the parity $P=\prod_r \tau_r^z$ to $P=1$ (this is easily found by rewriting $P$ in terms of the original $\sigma_j^z$ operators).

We now comment on the physical interpretation of the mapping to the Ising model outlined in this section. The Ising variables correspond to the vertex operators $A_+$ defined in the main text: $\tau_r^z=+1$ indicates the presence of an electric charge on site $r$, while no charge is present for $\tau_r^z=-1$. The (fixed) plaquettes represent a static background of magnetic fluxes, while the Wilson loops represent the fluxes around the two directions of the torus. The Hamiltonian in Eq. (\ref{eq:ising}) describes the dynamics of the charges in the static background of magnetic fluxes.

The lowest energy is obtained when no flux is present, i.e., when all the plaquettes are fixed to $B_{\triangle, \hexagon}=1$, and $W_{x}=W_y=1$. For $g>0$, this sector is mapped to the ferromagnetic Ising model on the kagome lattice, with $s_j=1$ for every $j$. The QSL phase in this sector is mapped to the paramagnetic phase in the ferromagnetic Ising model, which is known to exist up until $g_c/W=0.338678(4)$ \cite{Blote2002}. The correlation length of the ground state in this sector is associated with the mass of the electric excitation and vanishes for $g\rightarrow g_c$. On the other hand, for $g<0$, the no-flux sector is mapped to the antiferromagnetic Ising model on the kagome lattice. In this model, it is known that the paramagnetic phase extends along the whole $g<0$ line (as long as $W\neq0)$ \cite{moessner2001ising,Powalski2013}. This shows that the QSL phase exists in our model for $h=0, \lambda=0$ for any value of $g/W<0$. 

In the QSL phase, the lowest energy states in the sectors with $B_{\triangle, \hexagon}=1$ for every plaquette but different signs of Wilson loop operators have a gap $\propto \exp(-cL)$ (where $L$ is the shortest linear size of the system) with respect to the case $W_x=W_y=1$. Possible configurations of $s_j$ corresponding to these sectors are shown in Fig. \ref{fig:wegner}-b,c,d.

Another interesting quantity that we can compute exploiting this mapping is the mass of magnetic excitations. Because the product of all plaquettes is equal to $1$, we can only have an even number of magnetic excitations. Therefore, to measure the mass of a single excitation, we consider a sector with two excitations at very long distance (see Fig. \ref{fig:wegner}-e): we expect that their interaction energy becomes small for sufficiently long distance, such that the energy difference between the lowest-energy state in this sector and the ground state is equal to twice the mass of a magnetic excitation.

\label{sec:ising}
\section{Effective magnetic mass from perturbation theory}
\label{sec:perturb}
Each sector of the toric code Hamiltonian $H_0^{\text{TC}}$ can be mapped to a quantum Ising model on the kagome lattice with appropriate nearest-neighbor couplings, as shown above. We can apply perturbation theory to these Ising models in the small $g/W$ limit to estimate the ground-state energy for each sector. We note that perturbation theory is carried out within the regime of validity of the mapping we discussed above, and it is only needed to estimate energy splittings within allowed states in the spectrum.

At $g=0$, the ground state is the paramagnetic state with all  $\tau^z_r=-1$ and the energy is $E_0=-6WN_c$ in each sector. The first perturbative correction that depends on the sector is at sixth order, $E_6\propto -g^6 \sum_{\hexagon} B_{\hexagon}$, corresponding to flipping 6 Ising variables around a hexagonal plaquette. We see that, for both $g<0$ and $g>0$, the lowest ground state energy belongs to the sector with all $B_{\hexagon}=1$, and there is a finite gap (mass of magnetic excitations) separating this state from all the other sectors. Note that this mass is zero for $g=0$ and it is generated "dynamically" by the charge fluctuations, resulting in a gapped spin liquid ground state even in the absence of hexagonal plaquette terms.

\section{Additional numerical results}
\label{sec:num}

\subsection{Quantum Monte Carlo spectroscopy}

We provide here additional details on the QMC spectroscopy discussed in the main text. The toric code Hamiltonian in Eq. (2) commutes with the plaquette operators $B_\triangle$ and $B_{\hexagon}$, as well as with the non-contractible Wilson loop operators $W_i$, where $i = x,y$, defined as
\begin{equation}
    W_i = \prod_{j\in \gamma_i} \sigma_j^i,
\end{equation}
where $\gamma_i$ is a closed path crossing the periodic boundary conditions in direction $i$.
The spectrum of the toric code in the thermodynamic limit displays $4$ degenerate ground states, labeled by the signs of the non-contractible Wilson loops, and separated from the excited charged states by a gap, which at the perturbative level survives for $0<g/W<\widetilde g_c$, even in absence of plaquette terms. In our parameter regime, where magnetic charges are only allowed in hexagonal plaquettes, this gap corresponds at the leading order to corrections of order $g^6 \sum_{\hexagon} B_{\hexagon}$, as discussed in the previous section.

The verification of the existence of such a gap is a fundamental issue to assess the validity of the perturbative picture; however, a direct exact diagonalization (ED) analysis encounters strong difficulties due to the breaking of the ground-state degeneracy at finite system sizes. In particular, this phenomenon generates additional gaps of order $e^{-cL}$ between the $4$ formerly degenerate ground states, where $L$ is the shortest linear dimension of the system. These energy shifts may cause, for relatively small system sizes, the former ground states to rise in energy above the excited ones, preventing a clear spectral characterization of the model. In order to shed light on this issue, larger sizes than those accessible via ED need to be investigated.

To this end, we exploit the mapping between the superselection sectors of the toric code Hamiltonian in Eq.~(2) and inhomogeneous transverse-field Ising models, each characterized by the change in sign of a different subset of the nearest-neighbor couplings, discussed above.
Subsequently, we simulate the resulting quantum Ising models via Stochastic Series Expansion Monte Carlo \cite{Sandvik1991, Sandvik2003}, a state-of-the-art, unbiased method for the calculation of equilibrium properties of bosonic and unfrustrated quantum spin systems.

We compute the energies, on a cluster containing $N_{\mathrm{cells}} = 8 \times 8$ elementary cells, of the four quasi-degenerate ground states and of three charged states with increasing charge separation for $g/W = 0.32$. Each of these calculations corresponds to a ground-state analysis of a different quantum Ising model as discussed above; convergence to the ground-state regime has been verified in all cases by the compatibility within error of the results for inverse temperatures $\beta = 16,32$.

\subsection{Fidelity susceptibility}

\begin{figure}[h]
    \centering
    \includegraphics[width=0.4\linewidth]{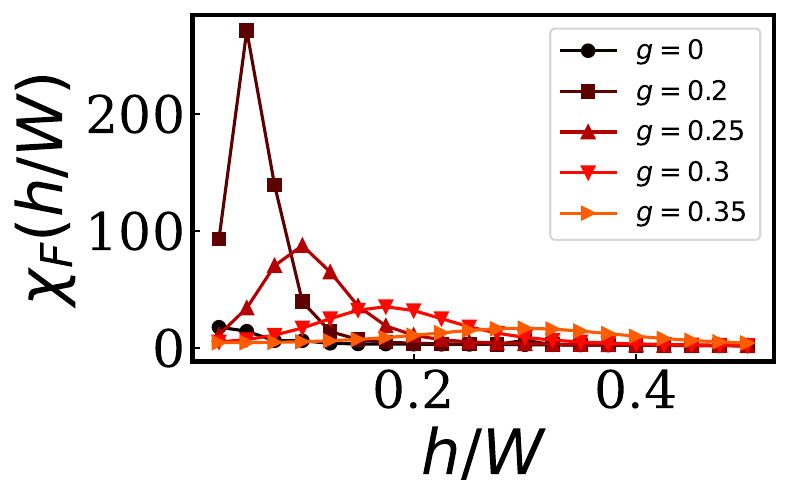}
    \includegraphics[width=0.4\linewidth]{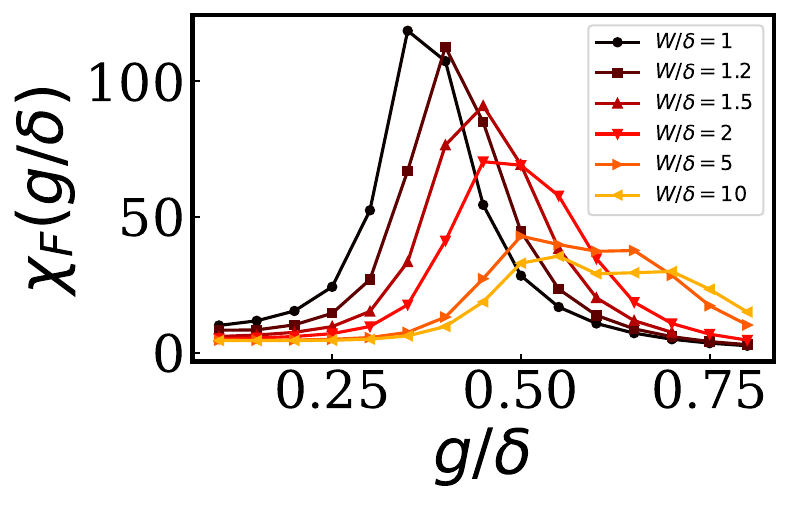}
    \includegraphics[width=0.4\linewidth]{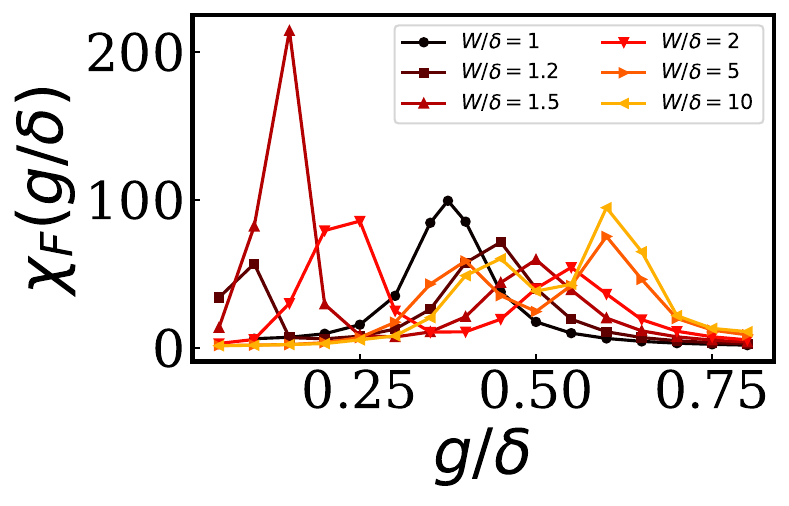}
    \caption{Fidelity susceptibility of the Hamiltonian $H^\text{TC}$ with (a) $\lambda=0$ as a function of the parameter $h$, (b) $g=h$ and (c) $h=0$ at various values of $W/\delta$.}
    \label{fig:fs}
\end{figure}

Fig. \ref{fig:fs}-a shows the fidelity susceptibility of the Hamiltonian $H^\text{TC}$ for $\lambda=0$ as a function of $h$. Fig. \ref{fig:fs}-b,c shows the
fidelity susceptibility at various values of $W/\delta$. We see that for $g=h$ the QSL phase, delimited by the two peaks in the fidelity susceptibility, appears only at large values of $W/\delta$, while for smaller $W/\delta$ a single peak is present, signaling a direct transition from a trivial phase to a valence bond solid (Fig. \ref{fig:fs}-b). In the case $h = 0$, we see that the QSL phase appears also at intermediate values of $W/\delta$ (see Fig. \ref{fig:fs}-c). The peaks of the fidelity susceptibility are taken to be the transition points shown in Fig. 2-b and Fig. 3-a,b in the main text.

\section{Alternative implementation along the $g=h$ line}
\label{sec:altexp}

The regime $h=g$ as realized in Ref.~\cite{Semeghini2021}, often referred to as frozen Rydberg gas, features interparticle interactions decaying as power law potentials: this induces substantial differences between $V_2, V_3$, and possibly next-neighbor interactions. This is because the native interactions between Rydberg atoms are induced dipole dipole ones, and only differ from that functional form at very short distances (where, however, a precise modelling of the dynamics as spin-1/2 model is not necessarily valid).

We propose here an alternative, double-dressing scheme that allows to independently tune $V_1/V_2$, and realize $V_2\simeq V_3$. We consider alkaline-earth like atoms trapped into deep optical potentials~\cite{cappellini2019coherent,norcia2019seconds,
madjarov2020high,wilson2022trapping}, so that atomic motion can be neglected. Within the low-energy manifold, we consider the electronic ground state $|g\rangle$ and the $^3$P$_0$ (or $^3$P$_2$) metastable state $|e\rangle$~\footnote{The hyperfine structure can be handled in different ways, depending on the laser polarizations.}\cite{schine2022long}. These two states represent the $n_j=0,1$ states, and are coupled by a clock laser, with Rabi frequency $\Omega_C$ and detuning $\delta_C$: these two quantities will determine $(h+g)$ and $\lambda$, respectively.
 
The metastable state is then off-resonantly coupled to two Rydberg states (see, e.g., Ref.~\cite{gil2014spin} for another proposal utilizing related settings), $|r_{1,2}\rangle$, with Rabi frequencies $\Omega_{1,2}$ and detunings $\delta_{1,2}$, such that $\delta_{1,2}\gg \Omega_{1,2}$, and $\Omega_1/\delta_1\gg \Omega_2/\delta_2$. Note that both of those couplings can be single-photon couplings to $s$-states, so that the resulting potentials are spatially isotropic. For both $^3$P$_0$ and $^3$P$_2$, they have already been experimentally engineered in Sr~\cite{madjarov2020high} and Yb bosonic isotopes~\cite{Takahashi2022}.

The resulting dressed potential is dominated by two contributions in perturbation theory \footnote{Note that, in the parameter regimes we work in, the resulting effective potential (while stll a combination of two soft-shoulder like contributions) is very different from the sum of two separate dressing potentials generated by distinct excitation patterns.}:
\begin{equation}\label{eq:doubldress}
V(r) = \frac{V^{(1)}}{1 + (r/r_1)^6} + \frac{V^{(12)}}{1 + (r/r_{12})^6}
\end{equation}
with $V^{(1)} = \Omega_1^4/8\delta_1^3$, $r_1 = ( C_6[11] / 2\delta_1)^{1/6}$, $V^{(12)} \propto \Omega_1^2 \Omega_2^2/\delta_1\delta_2^2$, $r_{12} \propto  ( C_6[12] / \delta_1)^{1/6}$, where $C_6[ij]$ are the van der Waals coefficients capturing the interactions between two atoms in the states $i,j$. Under the condition that $(C_6[11] / C_6[12])^{1/6}\simeq 2$, Eq.~\eqref{eq:doubldress} describes a potential that is very strong at short distances, with ratio $V_2/V_1\simeq (\frac{\Omega_2\delta_1}{\Omega_1\delta_2})^2 $, and then features a plateau that realizes approximately equal interactions for $V_2$ and $V_3$. 

\subsection{State preparation protocols}

In both implementations discussed, coupling to Rydberg states will necessarily lead to some amount of noise. In order to minimize the role of the latter, fast state preparation protocols are desirable. 

For the case of quantum spin liquids, the corresponding (diabatic) state preparation can take advantage of the relatively fast timescales that have been proven effective in preparing spin liquid states. In particular, in Ref.~\cite{giudici2022dynamical}, it has been shown that, even for moderate sizes of up to 70 spins, state preparation times of order of few tunneling processes shall suffice to realize very large overlaps with the target state. Since we do have access to an exactly soluble point, state preparation may further be shortened by starting from exact dimer coverings, following ideas already discussed in the context of strongly correlated fermions~\cite{trebst2006d}. 

An interesting, alternative path is to explore Floquet driving within the Rydberg manifold~\cite{scholl2022microwave}, that has already been demonstrated as a powerful tool to engineer long-ranged Heisenberg-like models~\cite{geier2021floquet}.

\bibliography{biblio}